# Visual and unplugged coding lessons with smart toys

Sara Capecchi
sara.capecchi@unito.it
CS Dept., University of Turin Turin, Italy

Cristina Gena
cristina.gena@unito.it
CS Dept., University of Turin Turin, Italy

Ilaria Lombardi
ilaria.lombardi@unito.it
CS Dept., University of Turin Turin, Italy

## ABSTRACT

Our Computer science k-12 education research group and the educational toy company Quercetti have been collaborating together to design and manufacture toys that help stimulate and consolidate so-called computational thinking. This approach is inspired by methods already consolidated in the literature and widespread worldwide such as the Bebras tasks and CS-Unplugged. This paper describes two smart toys, their design process, educational activities that can be proposed by teachers exploiting the two toys, the evaluation's results from some teachers, and finally feedback and reviews from buyers. The main activities proposed by these toys leverage visual coding through small colored physical items (e.g., pegs and balls) to deliver the unplugged activities to young users.

## 1 INTRODUCTION

Computer science is no longer an area of interest only for academics and professionals, but is fundamental for every citizen in order to cope with the ubiquity of information technology. It is of primary importance to understand the processes behind the digital world in which we are immersed and to learn the foundations of computer science by acquiring its basic conceptual tools. There is a general agreement on the fact that computer science should be part of general education from the earliest stages in order to develop computational and critical thinking skills. As outlined in the Proposal for a national Informatics curriculum in the Italian school [7] "*In the first stage (primary school) students should be encouraged to ask questions as well as to discover in their everyday life and to explore some basic ideas of Informatics. They can be engaged either in "unplugged" activities, i.e. without using digital technologies, possibly by drawing inspiration from the history of such ideas. Whatever the school level, the teaching of Informatics can, by its nature, be approached through active learning methods, teamwork and laboratory activities (including "unplugged" activities)*.

Computer Science Unplugged (CS Unplugged [1]) and Bebras tasks [4, 6] are a widely used collections of activities to introduce both children and adults to ideas and concepts from computer science, without having to use a digital device. With the variety of material appearing, the term "Unplugged" is currently used for computer science teaching activities that do not involve programming and the term refers to a general pedagogical approach. Some of the key principles that underpin the approach are: a sense of play for the student to explore even complex concepts, being highly kinesthetic and a constructivist approach.

Inspired by the above principles, our computer science k-12 education research group and the educational toy company Quercetti[1] have been collaborating together for several years to design and manufacture toys that help stimulate and consolidate so-called computational thinking. The paper describes two of these smart toys, the process that brought to their ideation, possible exercises teachers could make with them, the evaluation's results from some teachers that used them in classroom, and finally feedback and reviews from buyers. The proposed toys are based on visual coding through small colored physical items (e.g., pegs and balls) to be arranged in two-dimensional spaces, and this visual approach is fundamental to deliver these unplugged activities to young users.

This paper has been organized as follows: Section 2 presents the background and discusses the related work, Section 3 describes the ideation and the design of two smart toys and and some activities that can be done using them, Section 4 proposes two lessons that can be done in classroom with Pallino coding and some evaluation results, and finally Section 5 presents some feedback and reviews from buyers and concludes the paper.

## 2 BACKGROUND AND RELATED WORKS

The CS Unplugged material [2] is not intended to be used as a school curriculum, but as a form of pedagogy that has several potential benefits: the barrier of learning how to code, which can be seen as an insurmountable obstacle by some, is removed; it can be used in situations where computers are not available. Among the CS unplugged proposals that inspired our design of the Pallino coding there is *Marching Orders - Programming Languages*: in this activity about computer programming, learners follow instructions in a variety of ways in order to successfully draw figures. Through these exercises, learners experience some of the often frustrating aspects of programming. Moreover, by instructing each other on how to recreate a drawing, they can understand the importance of concise and explicit language that must be used in programming.

[1] https://www.quercettistore.com/
[2] https://www.csunplugged.org/en/

Bebras [3] is an international initiative aiming to promote computer science among school students at all ages. Participants are usually supervised by teachers who may integrate the Bebras challenge in their teaching activities. The tasks are fun, engaging and based on problems that computer scientists often meet and enjoy solving; they can be solved without prior knowledge but instead require logical thinking. They cover a wide variety of computer science topics and are designed to be usually solvable within 3 minutes. Thus, the tasks for the Bebras challenge have to concentrate on smaller learning items. There are many tasks related to the concepts covered by the proposed toys; those tasks can be used for learning assessment after the lessons we propose in Section 4.

The worldwide Code.org [9] (a nonprofit dedicated to expanding access to computer science in schools) also includes many unplugged lessons and activities in its courses. The ones that are most related to the proposed toys are: i) *Happy Maps, Happy loops, Graph paper based programming* in which students practice writing precise instructions as they work to translate instructions into the symbols provided and use symbols to instruct each other to color squares on graph paper. By "programming" one another to draw pictures, students get an opportunity to experience some of the core concepts of programming; ii) *Binary Bracelets and Binary images* in which students learn how information is represented in a way such that a computer can interpret and store it. When learning binary, they will have the opportunity to write codes and share them with peers as secret messages. This can then be related back to how computers read a program, translate it to binary, use the information in some way, then reply back in a way humans can understand.

Pixel art activities can also be used as a coding unplugged approach [3]. First using a *Declarative approach* the format used to represent (to code) images can be used as a rule to describe the image itself. Therefore the procedure to realize a picture is part of the rule used to describe it: the description of the image allows the executor to reproduce it, the image description is of *declarative type*. Instead, by using a *Imperative approach*, we can give a set of instructions describing what to do to draw an image. The description of the image and the steps needed to draw it are the same thing: for instance we can use instructions to program a robot both to move on a pixel matrix and to color the pixels it encounters.

In order to classify the educational activities proposed in Section 4 we used the *Proposal for a national Informatics in the Italian school* [7] which aim is to contribute to the development of Informatics education in the primary and secondary school in Italy. That proposal is the outcome of a long process, promoted by the Italian Informatics community, but that has also benefited from important contributions of pedagogists and experienced school teachers who took part in the discussion.

## 3 THE EDUCATIONAL SMART TOYS

Following our previous experience in educational robot co-design [5], [2], we co-designed the two educational smart toys with the team of Quercetti, a company producing educational toys for children from all over the world. The design of the two toys followed two different paths, having different requirements from the company. The Pallino coding's design originated from the idea of using already existing educational products as inspiration for possible CS unplugged activities; the Peg Code's design was guided by the idea of creating a new smart toy, possibly re-using pieces, and consequently molds, of already-existing toys. One of the company goals was to launch a new line of toys for CS unplugged activities, without electronic components in the toys (the company has never used them and would like to maintain this approach).

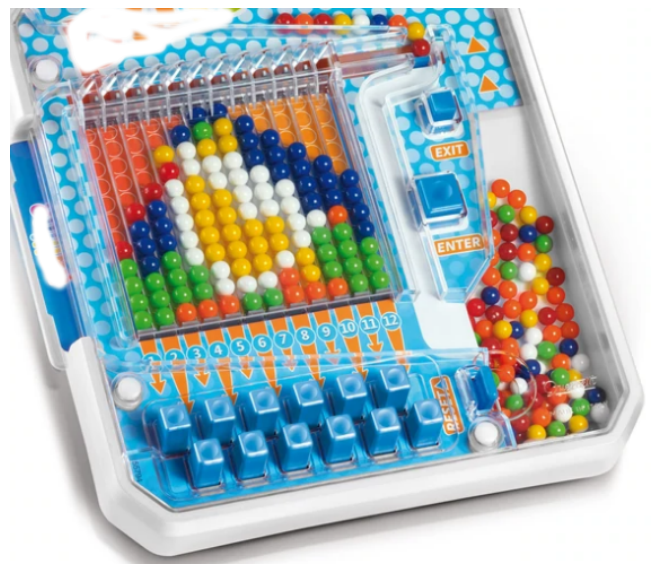

Figure 1: The original version of the Pallino coding

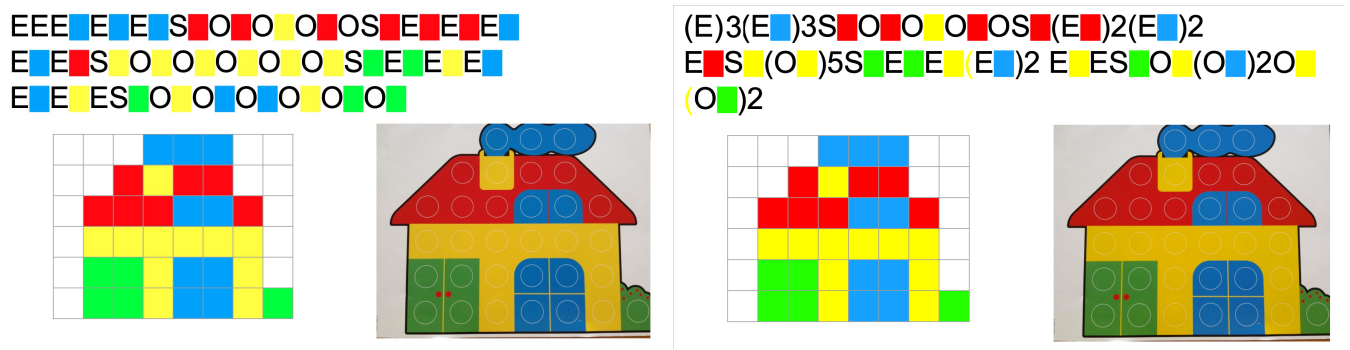

(a) (b)

Figure 2: 5-7 years old proposal (a), more than 7 year old proposal (b).

### 3.1 Pallino coding: the re-design

The Pallino coding's design originated from the idea of using an already existing toy as inspiration for CS unplugged activities, thus giving one of the old toys a new life. The company provided us a wide catalog of products that could be exploited for CS unplugged activities, and a set of toys to play with and then to propose new ideas. The design of the Pallino coding was developed through the following steps.

First, we selected a set of existing toys that we believed more suitable to introduce computer science concepts. On the basis of their features we proposed a set of unplugged activities, as the one depicted in Fig. 2a, whose purpose is to color the squares of a grid to create a drawing, following the instructions of a "program". The instructions reported in the first 3 lines of Fig. 2a have to be interpreted in order to reproduce the side drawing starting from the upper left square. The instructions have the following meaning: E (East) means *moving into the square to the right*, O (West) means *moving to the left square*, N (North) means *moving to the top square*, S (South) means *moving the bottom square*. The colors represent either the colors to be used to fill the squares on a sheet of paper or (looking at the available educational toys of Quercetti) how to combine the sequence of colored dots or pegs by inserting on a grid, to obtain the required drawing. We proposed these orienteeringbased activities because in the past we have found that children in the first years of primary school may have some orienteering problems and that coding can help them to improve, see [8]. We also made a second proposal for children older than 7 (see Fig. 2b), introducing the loop counter, namely the repetition of the same instruction: for instance the instruction (E)3 means move 3 times into the square to the right, (E blue)3 means for 3 times fill with color blue the square to the right, and so on. Then, we discussed the proposed activities with the company team (with members of both marketing and toy design units) with the purpose of selecting an existing toy that could be used both at home (in a setting with no educational purposes or trained adults) and at school for the above activities.

[3] www.bebras.org

On the basis of our proposals, the company team, among all the selected and suitable toys, decided for a renewal of the Pallino coding (see the old version in Figure 1) and for a re-design to best fit the unplugged coding activities described above. They also renamed it *Pallino coding*. The new version of the toy (described in the next Section) and additional material (cards, instructions) were co-designed in detail.

## 3.2 The Pallino coding

Pallino coding is a toy that stimulates children's creativity by reproducing colourful mosaics using "programmed" instructions. As they plan and make the mosaics, children take their first steps in the world of coding, i.e. programming[4]. There are two ways to play: the first is more creative and consists of making colourful mosaics following the mosaic patterns on the cards. The second lets children make up the pattern following the instructions on the "coding" side of the cards. When used in this way, Pallino coding may stimulate problem-solving strategies, such as algorithms, logic and breaking down problems. These are all elements which make up computational thinking, i.e. "thinking like a computer". Teaching children to think like this is becoming an important educational objective, in order to give them a head start in today's digital revolution.

With Pallino coding a children can do the following activities.

Reproduce a picture. By inserting a mosaic card (see Fig. 3) into the dedicated Pallino coding's slot, the children can reproduce and make the coloured balls line up with the same colours on the card. In Pallino coding (see Fig. 5), the screen is divided into 12 columns, and each column corresponds to a button. To launch the balls into a specific column, children push the launch button while holding down the corresponding column button. To remove the colours they don't need, they can press the "Exit" button.

If the children make a mistake, they can empty a single column by moving the "Reset" cursor while holding down the button for that column.

Read/Execute a program: Every mosaic card is two-sided. One side shows a mosaic card, (Fig. 3), on the other side shows a coding (or programming) card (see Fig. 4 and Fig. 5 upper right), which contains information to follow in order to reproduce a colourful mosaic. The coding card shows letters, which correspond with the rows, numbered starting from the bottom left, and numbers, which indicate how many balls of that colour should be placed in each row, from left to right.

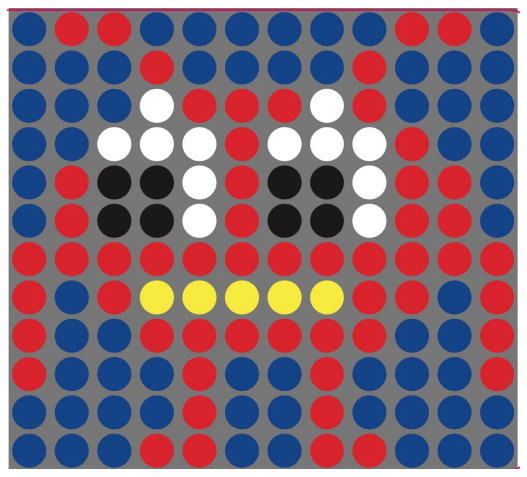

Figure 3: Reproduce a picture: the mosaic card

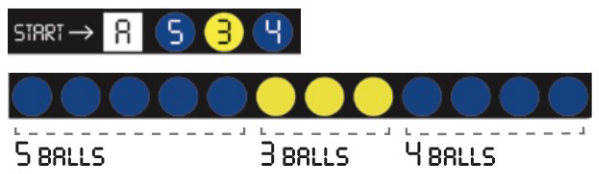

Figure 4: Reading a program: the coding card explained

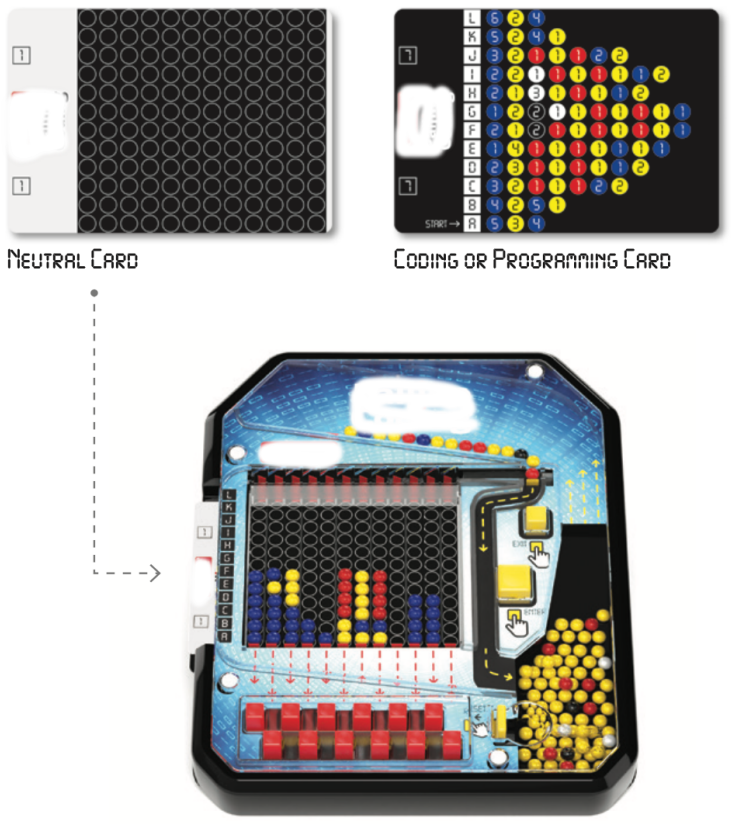

[4] Please note that in the paper we interchangeably use the terms coding and programming, see [10]

Figure 5: The Pallino coding in larger photo in the center. The neutral mosaic card (upper left) and the coding card (upper right)

Create/Reproduce a picture: The children can also make their own picture using a neutral mosaic card (see Fig. 5 upper left), and then reproduce it on the screen, in the same way explained in the case of the mosaic card of Fig. 3. In creating a mosaic the children can use the colours they want, paying attention to the quantity of balls of each colour contained in Pallino coding Coding: 64 yellow balls, 64 red balls, 64 blue balls, 16 white balls, 16 black balls.

Write/Execute a program: The children may also write their own program. In order to do is, they have to:

(1) Take a neutral mosaic card (see Fig. 5 upper left)

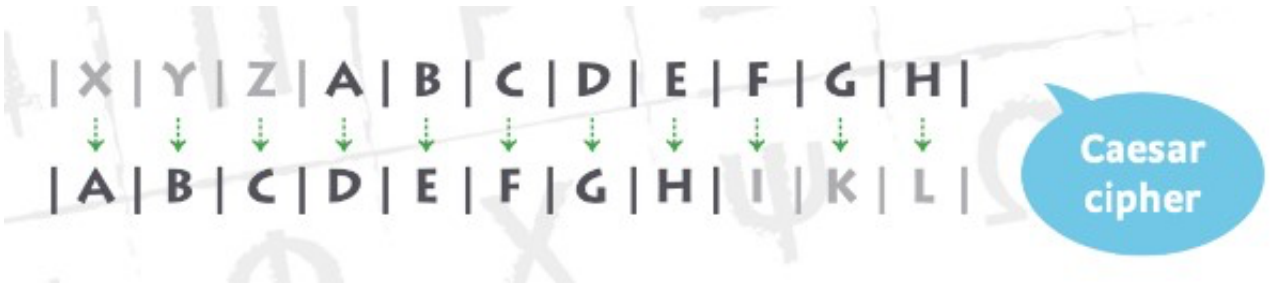

Figure 6: Caesar cipher

(2) On the neutral mosaic card, create a mosaic using the colours they want, paying attention to the quantity of balls, as explained above;

(3) Take a neutral coding card (the empty version of the coding card shown in Fig. 5 upper right), and write the instructions for reproducing the mosaic created at the step (2). The numbers written on the coloured circles correspond with the number of balls that need to be inserted on the same row from left to right;

(4) Start making the mosaic, following the instructions on the coding card. If the mosaic is the same as the one made on the first, then the program was correct.

This toy mode introduces the basic concepts of coding: in order to make a computer perform a task (a task planned by the us as "programrs") it must execute programs written in a language that has precise rules. Both the computer and the person who writes the program must know the rules of the language, i.e. which symbols it uses and what those symbols mean.

## 3.3 Peg Code: the co-design

As far as the design of the Peg Code is concerned, the idea was to create a new toy, possibly using existing pieces and molds. We started making some initial proposal related to i) orienteering activities through the use of checkerboard rugs to make either children move or objects move and ii) encryption toys exploiting secrete codes. This last proposal was the one selected for creating the new toy, the marketing department chose it for its potential appeal to children. We proceeded focusing on this idea, starting to propose possible ciphers/secrete codes to be used for an encryption toy. We introduced the company team to encryption starting to its historical roots and then we proposed them possible ciphers/secrete codes to be used for an encryption toy, as the Caesar cipher, the Pigpen cipher, and the Polybius cipher. We started the co-design by introducing the team to encryption and ciphers according to the following story-telling, which then became part of the toy guide. Encryption is a way to alter a message so that its original meaning is hidden. In order to decipher the message, the user needs a cipher, which is the system used to encrypt the transmitted message. Secret codes and ciphers are an important part of the science of communication security, i.e., cryptography. Throughout history, humanity has seen numerous events in which information needed to be transmitted from one person to another without being intercepted. It was fundamental during wars to send orders and to deploy troops, or in diplomacy to carry out negotiations before they were made official. In these cases, encrypted messages, ciphers and any means to make communication ‘secure’ were used.

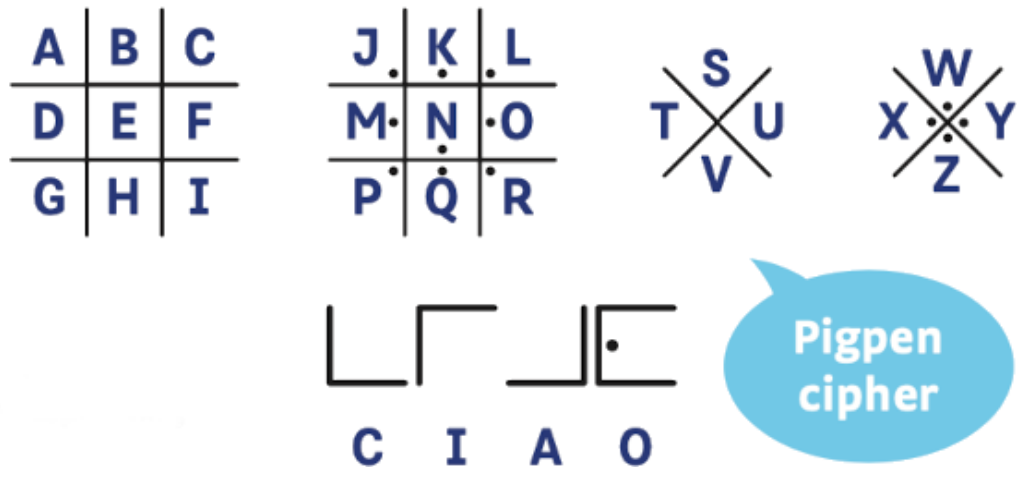

Figure 7: Pigpen cipher

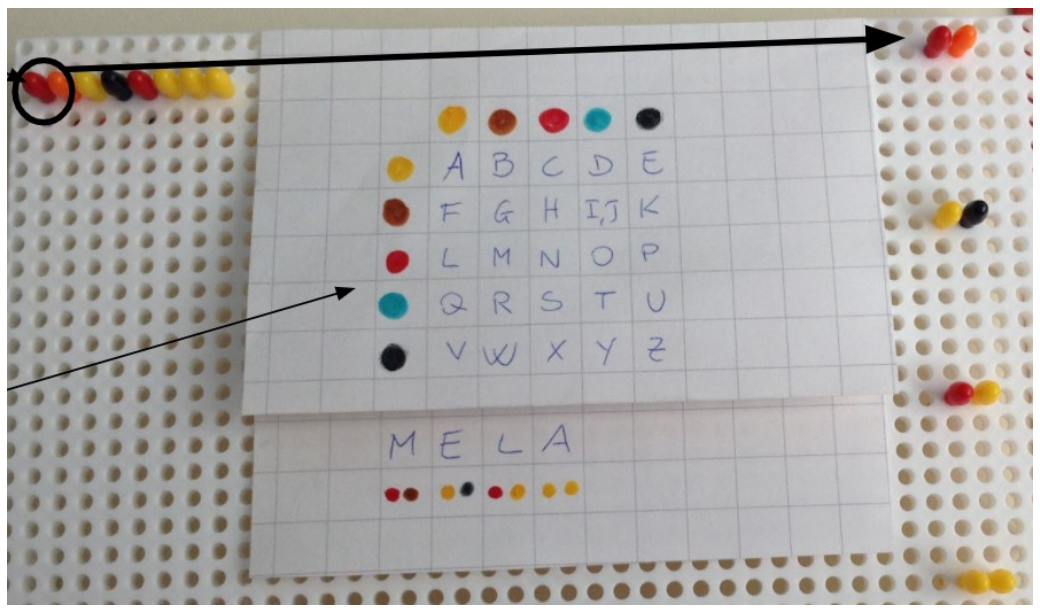

Figure 8: The Polybius cipher: first proposal

The Caesar cipher (see Fig. 6) is named after Julius Caesar, who used it to protect his secret messages. It was a very simple cipher, in which each letter was substituted with another letter. This type of cipher is quite weak because there are relatively few combinations, so it is easy to decipher based on the frequency of the letters used. In particular, Caesar used a shift of 3 positions (so the key was 3), like in the example shown in Figure 6. In order to create a toy based on this approach, we proposed a rotor for Caesar's cipher, which is a fairly common solution to make it work.

The Pigpen cipher (see Fig. 7), used by Freemasons since the 18th century, is a simple substitution cryptography system, in which letters are substituted with symbols. Each symbol corresponds with a letter. Once the user has figured out the system, she/he can decipher any message easily. For example, if we wanted to write the word 'CIAO', we would use the parts of the grid corresponding with the letters that we need, as shown in Figure 7. Among all the proposals about ciphers, the company team in the end chose to create the one based on the Polybius cipher, see Fig. 8, that used colored pegs instead of characters and numbers, as described in the following section.

## 3.4 The Peg Code cipher

The Peg Code cipher is based on our re-visitation of the the Polybius square and proposes a grid- and color-based encryption (see Fig. 9). The original Polybius square is characterized by following features: the square made up of a grid of 25 squares with 5 rows and 5 columns; the letters of the alphabet can be inserted onto the grid from left to right and from top to bottom; the rows and columns are numbered, and these numbers are the index, or "coordinates" of the letters in the message we wish to encode.

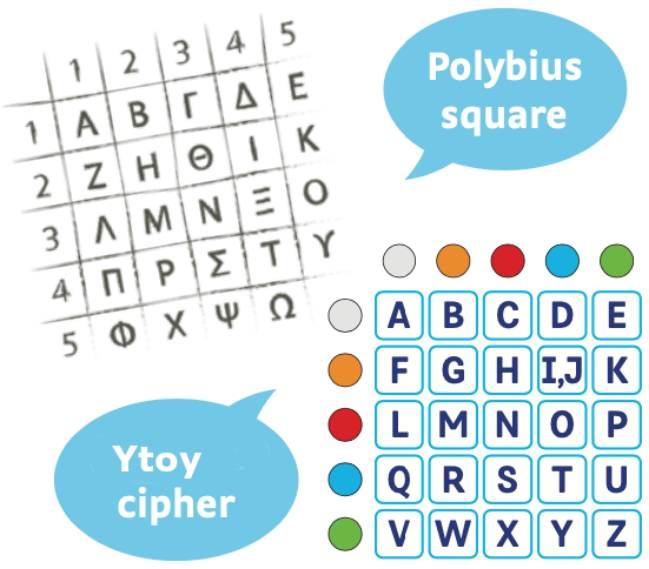

Figure 9: Polybious and Peg Code cipher

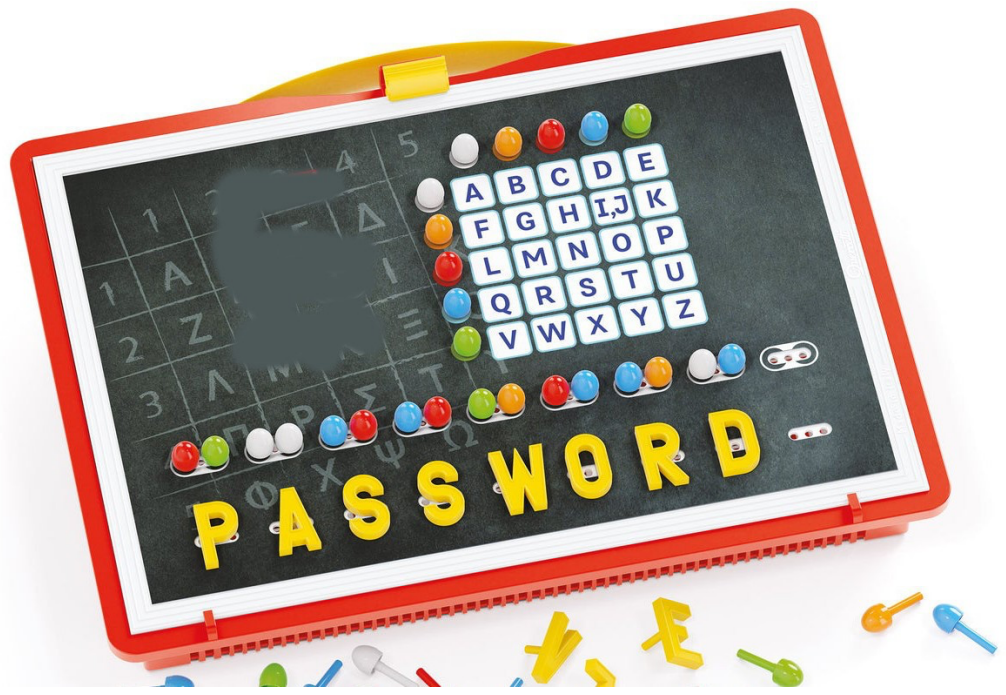

Figure 10: The Peg Code cipher

The YCode cipher is based on the same principle, but uses coloured pegs instead of numbers, so that a visual mapping is introduced, like shown in Fig. 9 and Fig. 10:

- white + white corresponds with the letter A;
- white + orange corresponds with the letter B; • white + red corresponds with the letter C, etc.

Thus in the Peg Code cipher, each letter is identified by two colors: the one of the row and the one of the column in which it is found. Children insert the pegs into the tablet using the right colour sequence and discover the code word by writing it with the letters provided. For example, in Fig. 11 we proposed to use the cipher backwards: the solution is written backwards under each sequence of letters, to make the decryption more tricky.

Other possible games with Peg Code are described in the following: (1) the player decodes the words encoded on the manual. This proposal is particularly suitable at the beginning, when children have to become familiar with the cipher; (2) one team encodes a word (possibly changing the combinations between colored pegs and letters, and thus creating a new version of the cipher) and the other team has to decode it (or a player from a team encodes and one from the same team decodes). The game goes on in this way, and the team who has decoded the most words wins; (3) two or more children who have Peg Code cipher on their own, share the same cipher and exchange secret messages that they alone can decode, because they alone share the same symmetric secret key. It can be seen again that the combination of colored pegs to encrypt a letter can be changed at will by the young user.

# 4 UNPLUGGED CODING ACTIVITIES FOR TEACHERS

After the Pallino coding's redesign was completed, we have created a sets of lessons proposing CS unplugged activities, which can be useful for teachers and parents. The lessons will soon be available on a dedicated web page. These two lessons have been tested in two classrooms, and the collected teachers' feedback is reported in Section 4.3. It should be noted that these lessons we propose could also be done without necessarily purchasing the toy, but simply using pre-filled (event manually set up by the teachers themselves) or neutral mosaics and coding cards.

## 4.1 Lesson 1: Run programs

Lesson description. In this activity the students may work individually, or in pairs: first they reproduce a drawing using Pallino coding following the mosaic card, then they reproduce the same drawing following the coding card. In this way they learn how an agent (a computer, a robot, a person) executes programs written in a language that has precise rules. Both the agent and the program author must know the rules of the language, that is, what symbols they use (syntax) and what they mean (semantics).
Features. Individual or couple activity. Age: from 7 years.
Duration: 2 hours
Aims. Learn to run programs by following their instructions. Understanding how an agent (namely who executes the program commands. It can be a robot, a computer or even a person) works.

Understand the rules of different programming languages. Understanding the need for programming structures such as loops. Material. A Pallino coding for each student or for each pair of students . No preparation is necessary as in this lesson only Pallino coding is used with the cards provided in the toy box.
The first activity. Students have to reproduce the drawing shown on a mosaic card (as the one in Fig. 3, the picture must be same for everyone). No explanation should be given at this stage, the teacher can show how to use the buttons, possibly with this video[5].

Then teacher and students discuss on the interpretation of the instructions on the mosaic card:

- What does a blue circle on the card mean? And a red one? The teacher explains that the color of the circle is part of the information that the agent must understand;
- Is location important? The teacher explains the meaning of position of the circle on the mosaic board;
- The teacher indicates where the ball should be placed. The location has two coordinates: a row (identified by a letter) and a column. The position of the circle is also part of the information that the agent must understand.

Thus an agent simply executes a program, that is a set of instructions; in this case the program describes to the agent the color of the ball that has to be placed in each position.

The second activity. To introduce the second activity, the teacher compares the coding card with the mosaic just created: she/he points out that there is also a colored circle on the coding card, which however has one more information: a number. The number expresses special information, i.e. how many balls of that color need to be placed. In other words, it indicates that students need to repeat the same operation several times. In programming languages, repetitions are called loops.

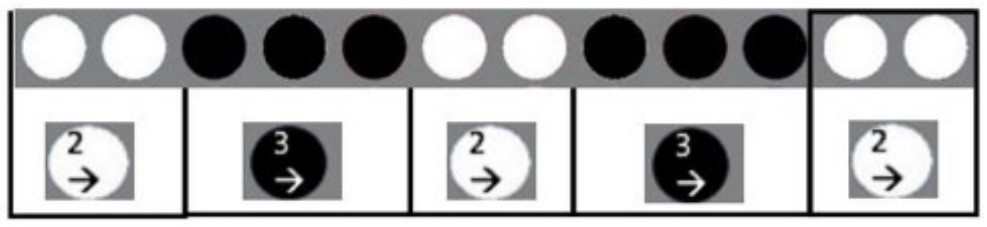

Figure 12: Coding with Pallino coding: the coding card explained

The correspondence between the colored circles of the coding card and the balls to be placed in the mosaic is explained in Fig. 12.

At this point, the students can reconstruct the mosaic, this time following the instructions on the coding card.

Final review and discussion. The teacher and the students reflect and discuss together on: *What difficulties did the students encounter?*

*It was easier to reproduce the mosaic with the indications of the mosaic card than with the ones of the coding card, or vice-versa? Did the students understood this programming language, that of the coding card? Did the students notice that, in the coding card, the sum of the numbers placed on each row is 12?*

Educational goals and objectives. This activity contributes to achieving the following goals and objectives, defined by *Proposal for a national Informatics in the Italian school [7]*: *Skills development milestones at the end of primary school*: i) understand that an algorithm describes a procedure that lends itself to being automated in a precise and unambiguous way; ii) understands how an algorithm can be expressed through a written program using a programming language; iii) read and write simple programs; iv) explain using logical reasoning why a program achieves its goals.

*Learning objectives at the end of the third grade of primary school*:

(*Algorithms scope*): i) recognize the algorithmic elements in routine operations of daily life; ii) understand that problems can be solved by breaking them down into smaller parts:

(*Programming area*): i) order the sequence of instructions correctly; ii) use cycles to synthetically express the repetition of the same action a number iii) set of times.

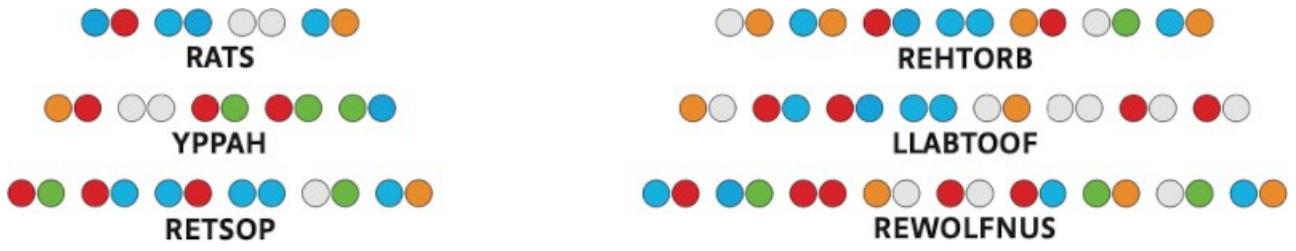

Figure 11: Peg Code cipher: an example

*Learning objectives at the end of the fifth grade of primary school*: (*Algorithms scope*): use logical reasoning to explain how some simple algorithms works.

## 4.2 Lesson 2: programming

Lesson description. In this activity the students work in pairs: in a first phase both create a colored mosaic on a paper (without showing it to the partner) and write a program to reproduce it; in the second phase, in turn, everyone performs the program written by their partner to create the mosaic with Pallino coding (or simply reproducing it on a mosaic card).

[5] https://www.youtube.com/watch?anomymousvideo

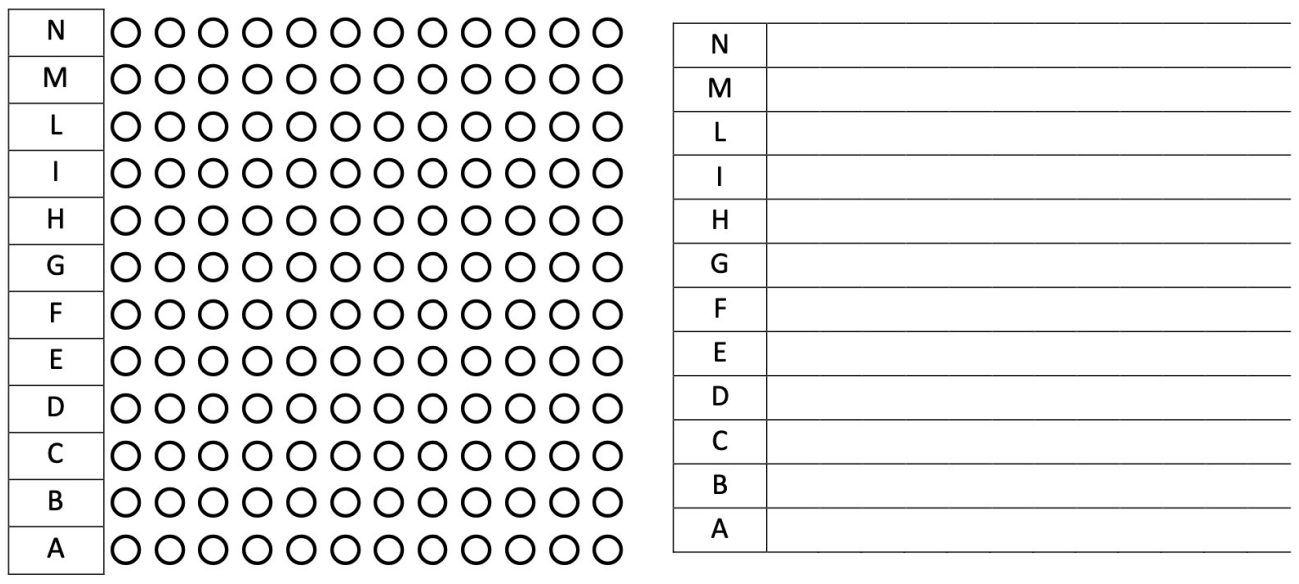

(a) (b)

Figure 13: Mosaic (a) and coding (b) cards

Features. Couple or group activity. Age: from 7 years. Duration: 2 hours.

Aims. Learn to think algorithms to solve real problems. Learn to program, that is, to translate algorithms into a programming language.

Material. An Pallino coding for each pair or group. Two mosaics and two coding cards for each pair, similar to the ones shown in Fig. 13. Markers or colored pencils with the same colors as the balls in Pallino coding.

Recap. The teacher summarizes the contents of the previous lesson with the students. She/he can start with some general questions:

*What did we do last time? What would you have liked to do? Since the last lesson, did you come up with any questions or reflections? What did you like most in the last lesson?*

Theactivity. The teacher starts by explaining that an agent (namely a robot, a computer, or a person) executes a program, that is a sets of instructions written in "its" language. Thus in order to program an agent students have to: *understand what are the actions that an agent can perform; know the agent's language, otherwise it is not possible to explain to it what it has to do (namely to program it).*

Expressing in a clear format the actions that has to be performed means developing an algorithm. The algorithm must then be transformed into a program, that is, a translation of the algorithm in the programming language that the agent understands. After the above explanation the teacher should remind students the meaning of the position and of the numbers of the circles with the following example, already used in the last lesson (see Fig. 12).

After that, the teacher can use the example shown in Figure 14. She/he shows the mosaic and asks students how to write the program to reproduce the mosaic using the coding card. At this point, some student may come up with alternative solutions, which have all to be discussed. For example, someone might suggest to proceed by columns instead of by rows. The right solution, the one expressed in the Pallino coding language, will be then written on the blackboard reproducing the coding card.

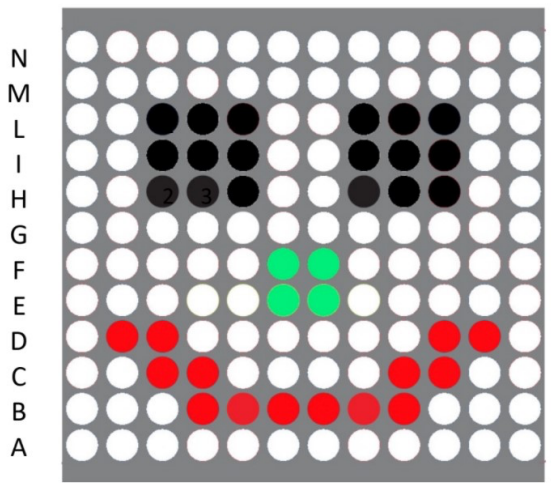

Figure 14: The mosaic to be described with instructions

At this point the students can move on to the first part of the activity: both members of the couple must make a new mosaic using the available colors, without showing it to the partner. They will be able to draw it directly on their mosaic card. When they both have finished their drawings, they have to write on their coding card the instructions to reproduce the mosaic in the language of Pallino coding, which they learned in the previous lesson. In the second phase of the activity, each student, in turn, will try to execute the program written by her/his partner, and only at the end will they check if they have both exactly reproduced the original mosaic. It may happen that for someone the result does not conform to the initial mosaic: in this case, the teacher urges the couple to look for and to fix the error, which may have been committed by whoever wrote the program or who executed it.

Final review and discussion. In the final part teacher and students discuss together, possibly answering the following questions:

- What difficulties did the students encounter?
- Did each student exactly reproduce the mosaic drawn by her/his partner?
- If not, did they manage to figure out what the problem was: was the program wrong or did they "execute" it wrong? Were they able to solve the problem?
- The teacher may stress on the fact that errors (that in jargon are called bugs) are common in a program. Thus programmers, in addition to writing programs, also have to look up for errors: this activity is called, in fact, *debugging*;
- Did the students remember the programming language of Pallino coding?
- Did the students remember what the numbers on the balls of the programming diagram mean? They indicate the repetition of the same ball several times, i.e. one loop;
- Have the students checked that the sum of the numbers placed on the same row was 12?

- It is a help to quickly notice some errors: surely if the sum is not 12 the students have made a mistake in writing the program, but it does not mean that there are no errors even if the sum is 12.

Activities classification. This activity contributes to achieving the following goals and objectives, defined by the *Proposal for a national Informatics in the Italian school [7]*:

*Skills development milestones at the end of primary school*: i) understand that an algorithm describes a procedure that lends itself to being automated in a precise and unambiguous way; ii) understand how an algorithm can be expressed through a written program using a programming language; iii) read and write simple programs; iv) explain using logical reasoning why a simple program achieves its goals.

*Learning objectives at the end of the third grade of primary school*: (*Algorithms scope*): i)recognize the algorithmic elements in routine operations of daily life; ii) understand that problems can be solved by breaking them down into smaller parts;

(*Programming area*): i) detect any failure in simple programs and take action to correct them; ii) order the sequence of instructions correctly; iii) use loops to synthetically express the repetition of the same action a number of times.

*Learning objectives at the end of the fifth grade of primary school*:

(*Algorithms scope*): i) use logical reasoning to explain the functioning of some simple algorithms; ii) solve problems by breaking them down into smaller parts; (*Programming area*): examine the behavior of simple programs also in order to correct them.

## 4.3 Some evaluation results

The two above described lessons have been tested in two classes, forth grade of primary school. The lessons were performed by the teachers exactly following the instructions above. After the activities, the teachers filled out a questionnaire to give us some feedback. They answered the following questions in a 4-points Likert scale:

- It was easy to understand the instructions (av. score:4);
- Was it easy to identify the steps to be carried out in sequence?
  (av. score: 4);
- The language of coding is understandable (av. score: 4);
- Rate the quality of the language (av. score: 4);
- Rate ease of reading (av. score: 3.5);
- Rate the pleasure and speed of the experience (av. score: 3);
- Give a general vote to the proposed activity (av. score: 3.5);

Then they were asked to freely answer the following questions (*we report the most useful answers immediately afterwards*):

- Where did you encounter the greatest difficulty during use?
  *In returning the discarded balls up.* (teacher #2);
- What would you improve and how within the proposed activity? *The use of an 8x8 or 10x10 matrix should be evaluated. After a while, children indulge in the toy without completing it, sometimes a little tired.* (teacher #2);
- Do you have any other observations? *The children understood the language of Pallino coding and did lesson 2 without difficulty (see Fig. 15). The exchange of instructions within the couple also allowed to broaden the discussion* (teacher #1).
  *Probably a removable side slider would allow to bring up the discarded balls without having to flip the toy. On several occasions, during this operation, the balls already inserted in the columns came out.* (teacher #2);

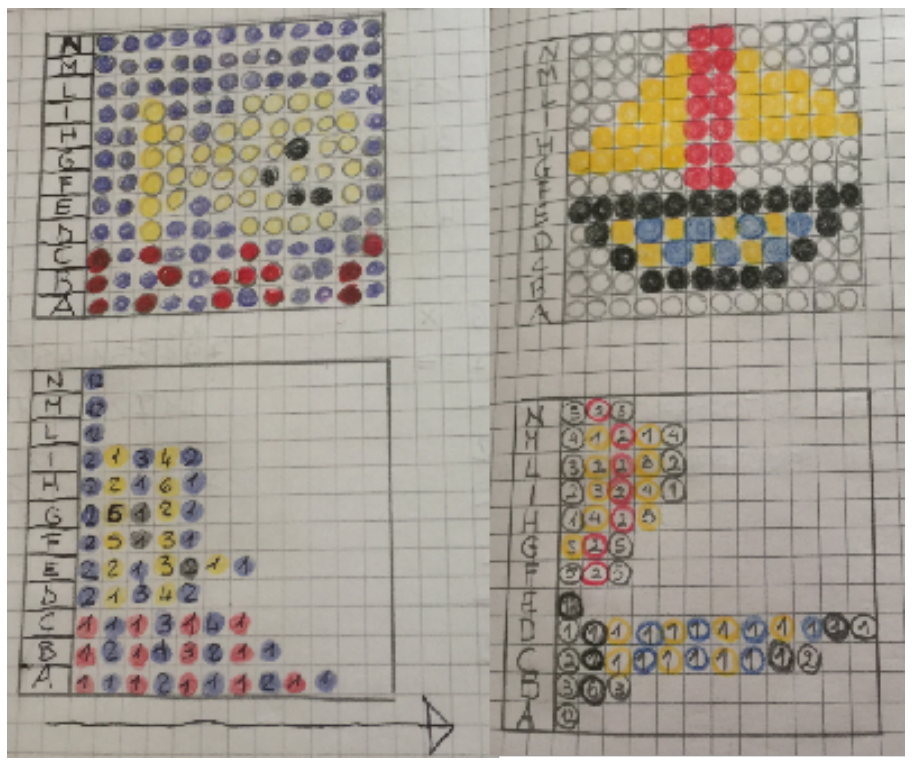

Figure 15: Two mosaic/coding cards by primary school children.

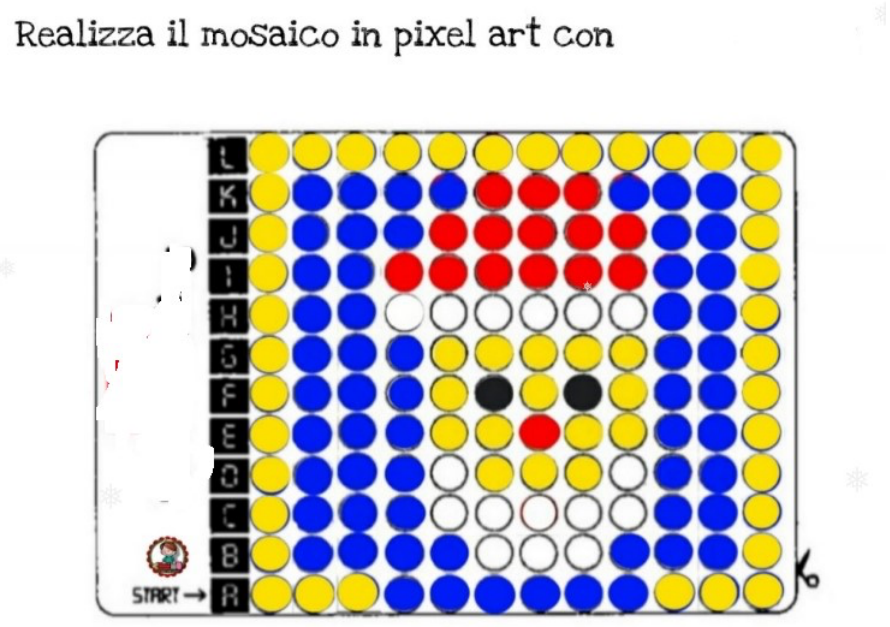

Figure 16: An educational activity proposed in an Italian blog for primary school teachers

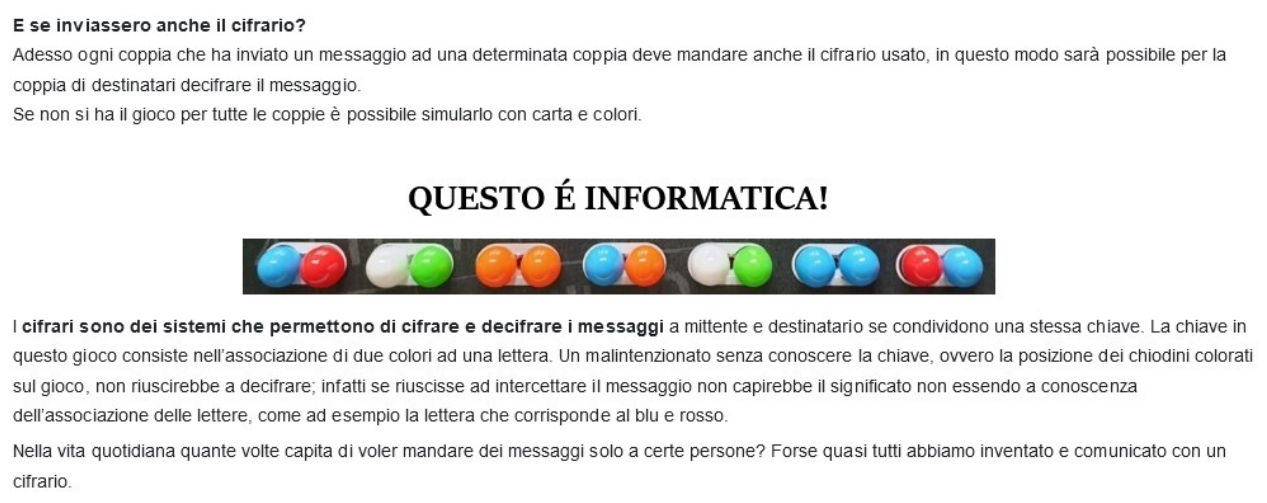

**E se inviassero anche il cifrario?**
Adesso ogni coppia che ha inviato un messaggio ad una determinata coppia deve mandare anche il cifrario usato, in questo modo sarà possibile per la coppia di destinatari decifrare il messaggio.
Se non si ha il gioco per tutte le coppie è possibile simularlo con carta e colori.

**QUESTO É INFORMATICA!**

I **cifrari sono dei sistemi che permettono di cifrare e decifrare i messaggi** a mittente e destinatario se condividono una stessa chiave. La chiave in questo gioco consiste nell'associazione di due colori ad una lettera. Un malintenzionato senza conoscere la chiave, ovvero la posizione dei chiodini colorati sul gioco, non riuscirebbe a decifrare; infatti se riuscisse ad intercettare il messaggio non capirebbe il significato non essendo a conoscenza dell'associazione delle lettere, come ad esempio la lettera che corrisponde al blu e rosso.

Nella vita quotidiana quante volte capita di voler mandare dei messaggi solo a certe persone? Forse quasi tutti abbiamo inventato e comunicato con un cifrario.

Figure 17: An educational activity proposed in an Italian educational project website

We have reported all comments back to the product designers in order to take them into account in future releases.

## 5 DISCUSSION AND CONCLUSION

Peg Code cipher was not evaluated with children before launch, because it was built and produced in the middle of the COVID19 pandemic (first half of 2020). On the other hand we got some feedbacks.

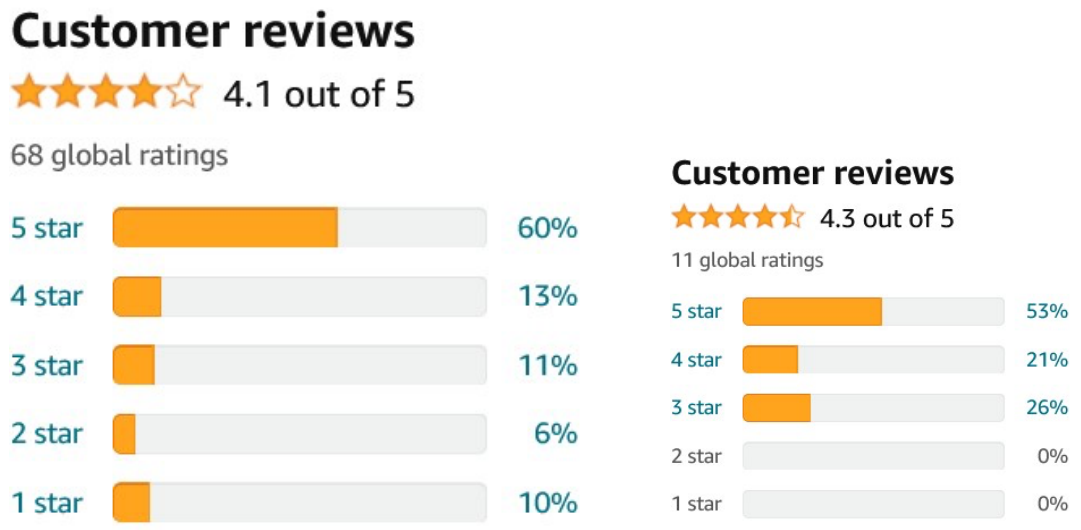

(a) Pallino coding ratings (b) Peg Code ratings Figure 18:

Amazon's ratings on Pallino coding and Peg Code

First, the two toys were used by teachers and researchers to plan educational activities/lessons for k-12 pupils (see Fig. 16 and 17)[6]. Finally, we report some data from the reviews on Amazon, which although not many are positive. Reviews of Pallino coding (see figure 18a) are generally positive. Negative feedbacks concerned mainly mechanics of the toys or damaged products due to delivery problems. Here we report some reviews:

- *Game that stimulates reasoning in children but are not satisfied with the mechanism. Very often the balls get stuck together and you risk losing the "job" done. To get the balls up, before placing them very often you risk getting out those already inserted in the track. Great idea but the mechanisms should be revised.*
- *Intelligent game to train children especially in the math field. It was a bit complicated to understand how it works, but just look at some tutorials on the web and it becomes easy to use. Useful for developing children's ingenuity.*
- *My 7-year-old son likes it a lot, it serves to improve concentration and intuition*

Feedbacks on Peg Code (see figure 18b) are positive too; some customers considered it too simple. Some reviews:

- *Only need to look for code in a 2d color matrix. No strategy or algorithm at all. If the design can involve some simple cipher algorithms that will be much interesting*
- *Nice educational, I bought this as a gift for my nephew. He totally loves it!*

We recognize that a limitation of our work lies in the fact that we have not been able to test the toys more extensively, especially Peg Code has only been informally tested in the designers' families before launch, because the children were all in lockdown at that time. Therefore, in the current context of our collaboration with Quercetti we are organizing activities with schools, where, among other things, we will carry out some activities with the two toys to test in the field, and with the children, their effectiveness and their weaknesses.

## 6 ACKNOWLEDGMENT

This work has been funded under the agreement between our university and Quercetti SpA. We would like to thank the Quercetti for the help and the inspiring exchange of idea during the overall collaboration.

## REFERENCES

[1] Tim Bell and Jan Vahrenhold. 2018. *CS Unplugged—How Is It Used, and Does It Work?* Springer International Publishing, Cham, 497–521. https://doi.org/10. 1007/978-3-319-98355-4_29

[2] Livio Bioglio, Sara Capecchi, Federico Peiretti, Dennis Sayed, Antonella Torasso, and Ruggero G. Pensa. 2019. A Social Network Simulation Game to Raise Awareness of Privacy Among School Children. *IEEE Trans. Learn. Technol.* 12, 4 (2019), 456–469. https://doi.org/10.1109/TLT.2018.2881193

[3] Alessandro Bogliolo. 2017. Pixel art, coding e immagini digitali. http://codemooc. org/pixel-art/

[4] Giuseppe Chiazzese, Marco Arrigo, Antonella Chifari, Violetta Lonati, and Crispino Tosto. 2019. Educational Robotics in Primary School: Measuring the Development of Computational Thinking Skills with the Bebras Tasks. *Informatics* 6, 4 (2019). https://doi.org/10.3390/informatics6040043

[5] Valerio Cietto, Cristina Gena, Ilaria Lombardi, Claudio Mattutino, and Chiara Vaudano. 2018. Co-designing with kids an educational robot. In *2018 IEEE Workshop on Advanced Robotics and its Social Impacts, ARSO 2018, Genova, Italy, September 27-29, 2018*. IEEE, 139–140. https://doi.org/10.1109/ARSO.2018.8625810

[6] Valentina Dagiene and Gerald Futschek. 2008. Bebras International Contest on˙ Informatics and Computer Literacy: Criteria for Good Tasks. In *Informatics Education - Supporting Computational Thinking*, Roland T. Mittermeir and Maciej M. Sysło (Eds.). Springer Berlin Heidelberg, Berlin, Heidelberg, 19–30.

[7] Luca Forlizzi, Michael Lodi, Violetta Lonati, Claudio Mirolo, Mattia Monga, Alberto Montresor, Anna Morpurgo, and Enrico Nardelli. 2018. A Core Informatics Curriculum for Italian Compulsory Education. In *Informatics in Schools. Fundamentals of Computer Science and Software Engineering*, Sergei N. Pozdniakov and Valentina Dagiene (Eds.). Springer International Publishing, Cham, 141–153.˙

[8] Cristina Gena, Claudio Mattutino, Gianluca Perosino, Massimo Trainito, Chiara Vaudano, and Davide Cellie. 2020. Design and Development of a Social, Educational and Affective Robot. In *2020 IEEE Conference on Evolving and Adaptive Intelligent Systems, EAIS 2020, Bari, Italy, May 27-29, 2020*. IEEE, 1–8. https: //doi.org/10.1109/EAIS48028.2020.9122778

[9] Filiz Kalelioğlu. 2015. A new way of teaching programming skills to K-12 students: Code. org. *Computers in Human Behavior* 52 (2015), 200–210.

[10] Enrico Nardelli (Ed.). 2020. *Coding e oltre: l'informatica nella scuola*. Lisciani Scuola Editore.

[6] We hide the links to the related web pages (an educational project website and a teacher blog) for the sake of anonymization.